# Spin-orbit coupling control of anisotropy, ground state and frustration in $5d^2$ Sr$_2$MgOsO$_6$

Ryan Morrow,[1*] Alice E. Taylor,[2*] D. J. Singh,[3] Jie Xiong,[1] Steven Rodan,[4] A. U. B. Wolter,[4] Sabine Wurmehl,[4,5] Bernd Büchner,[4,5] M. B. Stone,[2] A. I. Kolesnikov,[6] Adam A. Aczel,[2] A. D. Christianson,[2,7] Patrick M. Woodward[1]

[1] Department of Chemistry and Biochemistry, The Ohio State University, Columbus, Ohio 43210-1185, USA

[2] Quantum Condensed Matter Division, Oak Ridge National Laboratory, Oak Ridge, TN 37831, USA

[3] Department of Physics and Astronomy, University of Missouri, Columbia, MO 65211-7010, USA

[4] Leibniz Institute for Solid State and Materials Research Dresden IFW, D-01171 Dresden, Germany

[5] Institute for Solid State Physics, Technische Universität Dresden, D-01062 Dresden, Germany

[6] Chemical and Engineering Materials Division, Oak Ridge National Laboratory, Oak Ridge, TN 37831, USA

[7] Department of Physics and Astronomy, The University of Tennessee, Knoxville, TN 37996, USA

[*] Authors have made equal contributions

## Abstract

The influence of spin-orbit coupling (SOC) on the physical properties of the $5d^2$ system Sr$_2$MgOsO$_6$ is probed via a combination of magnetometry, specific heat measurements, elastic and inelastic neutron scattering, and density functional theory calculations. Although a significant degree of frustration is expected, we find that Sr$_2$MgOsO$_6$ orders in a type I antiferromagnetic structure at the remarkably high temperature of 108 K. The measurements presented allow for the first accurate quantification of the size of the magnetic moment in a $5d^2$ system of 0.60(2) $\mu_B$ – a significantly reduced moment from the expected value for such a system. Furthermore, significant anisotropy is identified via a spin excitation gap, and we confirm by first principles calculations that SOC not only provides the magnetocrystalline anisotropy, but also plays a crucial role in determining both the ground state magnetic order and the size of the local moment in this compound. Through comparison to $Sr_2ScOsO_6$, it is demonstrated that SOC-induced anisotropy has the ability to relieve frustration in $5d^2$ systems relative to their $5d^3$ counterparts, providing an explanation of the high $T_N$ found in $Sr_2MgOsO_6$.

## Introduction

There is a great deal of interest in materials with inherently frustrated magnetic exchange interactions and the resulting unusual magnetic ground states [1, 2]. One class of materials currently under investigation in this context is the double perovskites, formula $A_2BB'O_6$, where $A$ and $B$ are non-magnetic cations, and $B'$ is a magnetic $4d$ or $5d$ transition metal cation [3]. The resulting network of $B'$ cations forms a quasi-face-centered cubic (fcc) type lattice that is highly frustrated as each $B'$ cation has twelve nearest neighbor $B'$ cations [3]. Particularly complex magnetic states can arise in frustrated systems due to the presence of significant spin-orbit coupling (SOC) that typically accompanies cations of the $4d$ and $5d$ transition metals. It's imperative to understand the role SOC plays in such materials due to its role in lifting the orbital degeneracy and enhancing multipolar interactions, leading to rich phase diagrams [4-7].

In the case of double perovskites containing a single magnetic cation with a $d^3$ electronic configuration, the relatively large S = 3/2 spins are expected to have a quenched orbital contribution resulting in classical behavior [6]. While experimental results indicate that SOC is allowed in the $d^3$ configuration for $4d$ or $5d$ cations, the reduction in the magnetic moment due to

SOC is small relative to the effects of covalency, which is large for cations with high oxidation states [8, 9]. Double perovskites with $4d^3$ or $5d^3$ ions tend to exhibit frustration indexes ($|\Theta|/T_N$) ranging from 4 to 14 and adopt a type I antiferromagnetic structure upon ordering [8, 10-13], although incommensurate behavior has also been reported [14, 15].

In the case of the much smaller S = 1/2 $d^1$ configuration, SOC is expected to play a significant role and antiferromagnetic, ferromagnetic, and quadrupolar order have all been predicted as a result [5, 16]. These predictions appear to be in line with measurements of antiferromagnetic order in $Ba_2LiOsO_6$ and ferromagnetic order in $Ba_2NaOsO_6$ [17, 18]. However, there are additional examples such as spin glass $Sr_2MgReO_6$ [19] and spin singlet $Ba_2YMoO_6$ [20, 21] which do not easily fit into this framework. Investigation of magnetism due to the $d^4$ configuration has also begun, such as in $A_2BIrO_6$ (A = Sr, Ba; B = Sc, In, Y), where questions have arisen concerning the strength of SOC and the magnetism of the resulting ground state [22 - 25].

In this work, we investigate the influence of SOC on the magnetic state for the intermediate $5d^2$ S = 1 configuration, by studying the double perovskite $Sr_2MgOsO_6$ [26-28]. Theoretical work [6] has predicted a rich phase diagram with seven different phases/regions for the present $5d^2$ S = 1 scenario, however there has been difficulty in sorting known materials in this context. $Ba_2YReO_6$, $Sr_2YReO_6$, and $Ca_2MgOsO_6$ appear to be spin glasses, which have not been predicted [28-30], $La_2LiReO_6$ and $Sr_2InReO_6$ host non-predicted spin singlet states [29, 30], while μSR experiments indicate that cubic $Ba_2CaOsO_6$ orders antiferromagnetically, though with moments too small for detection in the reported neutron scattering experiment [31]. By combining magnetization, specific heat and neutron scattering measurements with first principles calculations for $Sr_2MgOsO_6$, we are able to provide insight into the nature of frustration and influence of SOC in this material.

$Sr_2MgOsO_6$ orders at 108 K with a type I antiferromagnetic structure, shown in Figure 1, a results that was previously predicted in a model considering the influence of SOC [6]. The $Os^{6+}$ moments of 0.60(2) $\mu_B$ are significantly reduced from the 2 $\mu_B$ spin only value expected for an S = 1 ion. Density functional theory (DFT) confirms that this substantial reduction in moment occurs through a combination of both covalency and SOC, and furthermore predicts that SOC-induced anisotropy is essential in the selection of the magnetic ground state. The presence of this anisotropy is experimentally confirmed by the observation of a spin gap in the magnetic excitation spectrum via inelastic neutron scattering. $Sr_2MgOsO_6$ is therefore a rare example of a compound where the Néel order, rather than just the anisotropy, is set by the spin-orbit interaction. Furthermore, we find that SOC-induced anisotropy is responsible for the reduced magnetic frustration in $Sr_2MgOsO_6$ relative to $d^3$ double perovskites, therefore explaining the enhanced $T_N$ in $Sr_2MgOsO_6$.

## **Experimental**

Powder samples of $Sr_2MgOsO_6$ were synthesized by grinding stoichiometric amounts of $SrO_2$, MgO, Os, and $OsO_2$ together using a mortar and pestle according to the following chemical equation:

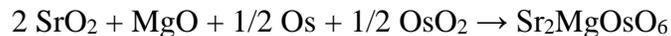

$$2\ SrO_2 + MgO + 1/2\ Os + 1/2\ OsO_2 \rightarrow Sr_2MgOsO_6$$

Ground mixtures of up to 3 g were contained in high-density alumina tubes and sealed in evacuated silica ampoules (approximate volume 40 mL with 3 mm thick walls) for heatings of 48 hours at 1000 °C in a box furnace located within a fumehood. This was followed by regrinding and identical reheating for an additional two cycles. Larger sample sizes were produced by synthesizing multiple aliquots which would be ground together in the intermittent grindings and redistributed for the subsequent heatings. Powdered $Sr_2MgWO_6$ samples were synthesized in air following the procedure outlined in the literature [32].

The temperature dependence of the magnetization of $Sr_2MgOsO_6$ powders was measured using a Quantum Design MPMS SQUID magnetometer. Data were collected over the temperature range 2.5 to 400 K under zero-field-cooled (ZFC) and field-cooled conditions (FC) in an applied field of 10 kOe. Powders were contained in gel capsules and mounted in straws for insertion into the device for measurement. An analogous data set was collected using an empty sample mount and subtracted from the temperature dependent magnetization data of $Sr_2MgOsO_6$ in order to remove the background response.

Powders of $Sr_2MgOsO_6$ and $Sr_2MgWO_6$ were cold pressed and sintered at their synthesis temperatures overnight (in an evacuated ampoule for $Sr_2MgOsO_6$) to prepare polycrystalline pellets. The specific heat measurements were conducted on the pellets mounted with Apiezon grease using a Quantum Design PPMS instrument using a relaxation technique.

Laboratory x-ray powder diffraction measurements were conducted at room temperature on a Bruker D8 Advance equipped with a Ge (111) monochromator and a Cu radiation source. Time of flight neutron powder diffraction (NPD) measurements were conducted on $Sr_2MgOsO_6$ at Oak Ridge National Laboratory's (ORNL) Spallation Neutron Source (SNS) on the POWGEN beamline [33] using a sample size of 1.359 g. Data were collected at 10, 50, and 300 K using the POWGEN Automatic Changer (PAC) environment. Separate data sets with the bank 2 and bank 7 chopper settings corresponding to respective d-spacing ranges of 0.2760–3.0906 Å and

2.2076–10.3019 Å were collected at each temperature. Data were analyzed using the Rietveld method as implemented in the GSAS EXPGUI software package [34, 35]. Additional NPD data were collected at High Flux Isotope Reactor (HFIR) facility at ORNL on the triple-axis spectrometer HB-1A using a sample size of 11 g. The sample was sealed under a He atmosphere into a cylindrical can made of aluminum with an inner diameter 0.6 cm. Data were collected at a constant wavelength of $\lambda=2.37$Å using collimation of 40′-40′-40′-80′. The data were analyzed using the Rietveld refinement suite FULLPROF [36], and the magnetic form factor for $Os^{6+}$ from Ref. [37] was assumed.

Inelastic neutron scattering experiments were performed on an 11 g sample of $Sr_2MgOsO_6$, and on a 16.5 g sample of $Sr_2ScOsO_6$ that was previously examined in Ref. [38]. Measurements were performed on the SEQUOIA chopper spectrometer at the Spallation Neutron Source (SNS) at Oak Ridge National Laboratory (ORNL). The samples were sealed in aluminum cans, and an identical empty Al can was measured as a background. A closed-cycle refrigerator was used to reach temperatures between 6 K and 125 K, and measurements were performed using an incident neutron energy 20 meV. Empty-can measurements were subtracted from the data sets, which were then normalized by a factor $m_{f.u.}/m_s$, where $m_{f.u.}$ is the formula unit mass and $m_s$ is the sample mass for each of $Sr_2MgOsO_6$ and $Sr_2ScOsO_6$. The presented magnetic scattering intensity is therefore per Os ion.

First principles calculations were performed using the generalized gradient approximation (GGA) of Perdew, Burke and Ernzerhof (PBE) [39] with the general potential linearized augmented planewave (LAPW) method [40] as implemented in the WIEN2k code [41]. We used the experimental 10 K crystal structure and highly converged basis sets, including local orbitals with LAPW sphere radii of 2.0 bohr for the metal atoms and 1.55 bohr for O. We did calculations both with the PBE-GGA itself and with an additional Coulomb repulsion parameter U=3 eV in the PBE+U approach with the fully localized limit double counting. The key difference between these treatments is that metallic behavior is predicted with PBE while an insulating gap is obtained with U=3 eV, consistent with experimental reports for the resistivity of polycrystalline samples [28]. We focus on results that do not depend on U.

## Results

$Sr_2MgOsO_6$ crystallizes in the tetragonal $I4/m$ space group as previously reported [27, 28] and shown in Figure 1, which is common to a number of other $Sr_2BOsO_6$ compositions [42-46]. The $I4/m$ space group is associated with the $a^0a^0c^-$ Glazer tilt system, where out of phase tilting occurs about the $c$-axis [47]. Rietveld refinement of the x-ray powder diffraction data did not indicate any disorder between Mg and Os cations, a typical result considering the charge difference of 4+ between the anticipated oxidation states of $Mg^{2+}$ and $Os^{6+}$ [3]. Refinements also did not indicate any loss of Os during synthesis within experimental error.

The results of the refinement of neutron powder diffraction data on $Sr_2MgOsO_6$ from POWGEN are given in Table 1, and the refined pattern at 10 K is given in Figure 2. The average Os−O bond length is typical of recent results for octahedrally coordinated $Os^{6+}$ in the double perovskite structure [42, 46, 48]. Both the $MgO_6$ and $OsO_6$ octahedra are slightly elongated, with two slightly longer and four slightly shorter M−O bonds. For the $d^2$ cation $Os^{6+}$, this is the expected Jahn-Teller distortion. The unit cell is tetragonally distorted as compared to the cubic cell, with a $c$-axis which is greater than $\sqrt{2}$ of the $a$-axis. The tetragonal distortion is enhanced at lower temperatures, with a $c/\sqrt{2}a$ ratio of 1.0068 at 300 K and 1.0206 at 10 K. The distortion manifests as a combination of enhanced tetragonal elongation of the octahedra and octahedral tilting as evidenced by a reduced Mg−O−Os bond angle. No structural phase transition or change in symmetry occurs within the temperature range studied. The short Os−O bond lengths compared to the $d^3$ osmates reflect the contraction and increased covalency that can be anticipated as the oxidation state is increased.

The temperature dependence of the magnetization of $Sr_2MgOsO_6$ is given in Figure 3a. A clear cusp corresponding to an antiferromagnetic transition occurs with a maximum at 108 K in both the FC and ZFC data sets, in approximate agreement with previous reports [27, 28], while the Fisher heat capacity, $d(\chi T)/dT$, shown in Figure 3b, indicates a maximum value at 102 K – a lower $T_N$ from this approach is typical of double perovskites [31]. A divergence of the FC and ZFC data at 15 K is noted in Figure 3a which is absent in previous reports [27, 28] despite being present in numerous independent samples and measurements by the present authors. A Curie-Weiss fit was conducted in the temperature range 250 to 400 K, which resulted in an effective moment of 1.88 $\mu_B$, in close agreement with reported values, and a Weiss constant of $\Theta = -269$ K, in approximate agreement with reported values [27, 28]. The effective paramagnetic moment is substantially reduced from the theoretical spin-only result of 2.83 $\mu_B$ for S = 1, indicating that the influence of spin-orbit coupling is significant for $Os^{6+}$. The ratio between the Weiss constant and ordering temperature $T_N$ yields a relatively low frustration index, ($|\Theta|/T_N$), of 2.5.

The specific heat of $Sr_2MgOsO_6$ is shown in Figure 4, with a clear second-order type anomaly positioned at 108 K indicating that the transition is likely due to long-range antiferromagnetic order. An isostructural nonmagnetic material with similar mass, $Sr_2MgWO_6$, was measured to approximate the nonmagnetic lattice contributions to the specific heat. The solid line, also shown in Fig. 4(a), represents the specific heat data of $Sr_2MgWO_6$ scaled by mass. The difference of these two data sets, plotted as $C_{mag}/T$ and shown in Figure 4b, corresponds to the magnetic component of the specific heat in $Sr_2MgOsO_6$. Clearly, there is a large peak at the antiferromagnetic transition, but there is also a significant tail up to temperatures much higher than $T_N$, indicating persistent magnetic fluctuations. The magnetic specific heat $C_{mag}/T$ is integrated to obtain the magnetic entropy, plotted against the right axis in Figure 4b. Analysis of the magnetic entropy over the entire temperature range results in $S_{mag} \sim 11.2$ J/mol K, which is in between the theoretical values for a simple L-S scheme for a $d^2$ cation with an expected total spin J = 2 ($S_{mag}$ = 13.38 J/mol K) and a spin-only scenario with S = 1 ($S_{mag}$ = 9.134 J/mol K). These results clearly contrast from a similar analysis recently conducted on $3d^8$ S = 1 $Sr_2NiWO_6$ [49], where a spin-only analysis in a nominally orbitally quenched material resulted in good agreement with the experimentally determined magnetic entropy. Therefore, we conclude that the magnetic entropy of $Sr_2MgOsO_6$ is significantly impacted by the orbital contribution to the magnetic moment in this compound.

In order to obtain a microscopic insight into the ordered magnetic structure, we examined the low $Q$ region of the neutron powder diffraction data collected on the POWGEN instrument, shown as the inset of Figure 2. However, no apparent magnetic reflections were observed arising below the ordering temperature. The specific location of the anticipated peaks associated with the common type I antiferromagnetic order are highlighted in the inset. This is similar to the case of $5d^2$ $Ba_2CaOsO_6$, where NPD data from the C2 diffractometer at the Canadian Neutron Beam Centre at Chalk River National Laboratories did not yield any observable magnetic reflections despite evidence of long range magnetic order from muon spin relaxation experiments [31]. In that study, it was determined that the ordered moments must be less than an estimated detection limit of 0.7 $\mu_B$ per $Os^{6+}$.

In order to search for the presence of weak magnetic reflections, additional neutron powder diffraction data was collected on the HB-1A beamline using an 11 g sample, nearly 10 times the mass measured on POWGEN, and with the sample contained in an aluminum can to minimize incoherent scattering. HB-1A was utilized for this particular investigation because of its excellent signal-to-noise ratio, arising from the combined use of a double-bounce monochromator and an analyzer. Below $T_N$, two magnetic reflections previously anticipated due to type I antiferromagnetic order were observed, see Figure 5. The diffraction pattern was analyzed using constant structural values as determined from POWGEN, but varying the background and instrument-dependent parameters, resulting in a good fit to the data, Figure 5a. Extra peaks are visible which are due to the aluminum sample can scattering and a small, unidentified impurity phase is visible in this sample, as indicated in Figure 5, which is present at all temperatures. The magnetic structure refinement yielded $Os^{6+}$ moments of 0.60(2) $\mu_B$, which are aligned within the *a-b* plane. The resulting value is just below the proposed detection limit from the case of $Ba_2CaOsO_6$ [31]. The temperature dependence of the $Q(001) = 0.78$ Å$^{-1}$ peak is shown in Fig. 5c. A power-law curve was fit to the data to extract $T_N$, and confirms the $T_N$ = 108(2) K transition temperature associated with this magnetic peak. This is a remarkably high Néel temperature for a double perovskite with B = Mg, since the $Os^{6+}$ ions are on a quasi-fcc lattice and are separated by more than 5.5 Å.

For our DFT calculations, we considered different magnetic orders including ferromagnetic, the observed type I order, and checkerboard antiferromagnetic order in the basal *a-b* plane of the tetragonal cell. We find that PBE-GGA calculations without SOC predict an incorrect magnetic order, specifically a ferromagnetic ground state. Only when spin-orbit coupling is included do we obtain the correct type I order as the lowest energy state. This conclusion is robust, as when spin orbit coupling is included we obtain type I order independent of the moment direction and for both PBE and PBE+U calculations. It follows that $Sr_2MgOsO_6$ is a rare example of a material where the Néel order itself, rather than just the anisotropy is set by the spin-orbit interaction – a reflection of the strong SOC.

Also for both PBE and PBE+U calculations with SOC we find the lowest energy spin direction to be along the tetragonal <100> direction in the tetragonal *I*4/*m* cell. The anisotropy is sizable and increases with U. For the PBE calculations we find that the <110> direction is disfavored by 1 meV/Os, while the <001> direction is disfavored by 5 meV/Os. The easy axis and plane do not depend on U. This anisotropy and the symmetry breaking due to the tetragonal lattice no doubt partly explain the relatively high ordering temperature on a dilute fcc-like lattice. The other needed ingredient in obtaining the ordering is the intersite exchange interaction. The value of this coupling depends on the choice of U, and as such we cannot directly predict the precise magnitude of this coupling. However, as seen from the energy differences, regardless of the choice of U the correct ground state is predicted, i.e. there is sufficient intersite exchange coupling. We find that the ferromagnetic ordered state is 31 meV above the ground state in the PBE+U calculation and 168 meV above the ground state without U.

Turning to the moment size, based on integration within the LAPW spheres (2.0 bohr for Os) we obtain moments that are strongly reduced from the nominal values due both to covalency and SOC. The total Os moment in the PBE calculation is 0.48 $\mu_B$ consisting of a spin moment of 0.77 $\mu_B$ and an orbital moment of −0.29 $\mu_B$. In the probably more realistic PBE+U calculation we obtain 0.57 $\mu_B$, from a spin moment of 1.07 $\mu_B$ and an orbital moment of −0.50 $\mu_B$, in close agreement with the experimental result of 0.60(2) $\mu_B$ from NPD. We also find sizable moments on the O ions, which is a result of strong covalency. This can be seen in the density of states, shown for the ground state with PBE+U in Fig. 6. The O 2p bands extend from −7.4 eV to −1.1 eV relative to the valence band maximum. The region from the bottom to −5.1 eV comprise O 2p – Os $e_g$ σ bonding states, and one can see very strong Os character in this region reflecting the covalency. The Os $t_{2g}$ states, which are the active orbitals here, extend from −0.5 eV to +1.8 eV and are split by U to give a gap of 0.22 eV. The $t_{2g}$ band width is similar in the PBE calculation. The Os $e_g$ states extend from 4.3 eV to 6.3 eV. The large crystal field splitting is another reflection of very strong covalency.

A consequence of this covalency is the presence of sizable moments on the O sites. In the double perovskite structure each O has only one Os nearest neighbor. The O moments are parallel to those of the neighboring Os regardless of the treatment (PBE or PBE+U), the magnetic order or the inclusion of SOC. While $Os^{6+}$ $d$ orbitals can be regarded as largely inside a 2 bohr LAPW sphere, this is not the case for $O^{2-}$ $p$ orbitals with a 1.55 bohr sphere. In order to estimate the O contribution, we turn to the ferromagnetic case. In the PBE+U calculation the total spin moment in the unit cell is 1.93 $\mu_B$ per formula unit, while the Os spin moment is only 1.14 $\mu_B$. Thus ~40% of the spin moment is distributed over the six neighboring O ions. This moment will be active in susceptibility fits, but not in refinements of neutron diffraction data as the moment is spread across neighboring oxide ions and bonds. The O moments provide an explanation for the sizable intersite exchange. Although the double perovskite lattice can be regarded from a Zintl perspective as touching $(OsO_6)^{6-}$ anions held together by interstitial $Mg^{2+}$ and $Sr^{2+}$ cations, the fact that the O atoms on the exterior of these contacting polyanions carry sizable moments provides a mechanism for the intersite exchange.

Having shown that SOC has a significant influence on the moment size observed by neutron scattering, we anticipate that SOC may have a major effect on the magnetic dynamics in

Sr$_2$MgOsO$_6$. Confirmation of this can be found by examining the inelastic neutron scattering spectra of both Sr$_2$MgOsO$_6$ and Sr$_2$ScOsO$_6$ shown in Fig. 7. We compare Sr$_2$MgOsO$_6$ to Sr$_2$ScOsO$_6$ because Sr$_2$ScOsO$_6$ also shows high-$T_N$ type I AFM order with $T_N$ = 92 K, but has Os$^{5+}$ $5d^3$ ions which are expected to show significantly reduced SOC due to a S=3/2 state. In both materials we observe scattering emanating from the type I antiferromagnetic wavevector at $Q \approx 0.8$ Å$^{-1}$, Fig. 7. We identify the development of a spin gap in both materials at low temperatures—compare Fig. 7(a) to Fig. 7(c) for Sr$_2$MgOsO$_6$ and Fig. 7(b) to Fig. 7(d) for Sr$_2$ScOsO$_6$. The Sr$_2$MgOsO$_6$ spectrum at 6 K is remarkably similar to that of Sr$_2$ScOsO$_6$, in which the gap has been extensively characterized [38], see Figs.7(a) and (b), respectively. This suggests that the physical mechanisms controlling each system are more similar than previously predicted [6], with SOC having influence in both materials.

The similar size of the gaps in Sr$_2$MgOsO$_6$ and Sr$_2$ScOsO$_6$ does, however, support a picture of SOC having stronger influence in Sr$_2$MgOsO$_6$. The microscopic mechanism by which SOC typically produces the spin gap is associated with either exchange anisotropy or single-ion anisotropy (or a combination) [10, 15, 38, 50]. For either mechanism, for a fixed strength of SOC the magnitude of the gap observed by neutron scattering scales with the magnetic moment size. Therefore, as Sr$_2$MgOsO$_6$ has a smaller magnetic moment, the similarity of observed gap to that in Sr$_2$ScOsO$_6$ demonstrates that SOC is stronger in Sr$_2$MgOsO$_6$ resulting in a comparable gap.

Despite the similarity in the spectra at 6 K, above $T_N$ the intensity of the observed scattering is very different between the two compounds, see Figs. 7(c) and (d). To examine the temperature dependence further, we present in Fig 8(a) the integrated intensity of the scattering for the range $0.7 < Q < 1$ Å$^{-1}$ and $3 < E < 12$ meV. For both materials the integrated intensity in this region increases with temperature, because of both the modification of the scattering due to the closing of the gap and the Bose thermal population factor. The integrated intensity per Os ion for each of the samples is similar at low temperatures, with a slightly higher value for $Sr_2ScOsO_6$, as expected for the larger spin system. However, as the magnetic transition temperatures are approached the integrated intensities diverge dramatically. This difference in fluctuation intensity above $T_N$ is indicative of the level of frustration in each system - a strong signal implies strong correlations despite the absence of long range magnetic order.

In Fig. 8(b) we present the temperature dependence of the scattering at very low energies, i.e. within the gap at low temperatures. The data is converted to $\chi''(T)$ for the fixed range $0.7 < Q < 1$ $\text{Å}^{-1}$ and $3 < E < 5$ meV following the method described in Ref. [10], in which the lowest temperature data set has been subtracted as a background and a Bose factor correction has been applied. This confirms that the reduction in the scattering at low energies is far beyond what would be expected due to thermal population, thereby confirming the opening of a gap at low temperature.

## Discussion

A phase diagram depicting seven potential magnetic ground states, including three potential antiferromagnetic configurations, has been proposed for double perovskites with a single magnetic $5d^2$ cation [6]. Through a combination of reduced paramagnetic effective moment, an enhanced magnetic entropy from a spin-only scenario, a substantially reduced moment refined from neutron diffraction, and a significant spin-excitation gap observed in neutron spectroscopy, we have unequivocally shown that SOC plays a major role in the magnetic behavior of the $Os^{6+}$ $5d^2$ cation in $Sr_2MgOsO_6$. Despite this, we have shown that the ground state remains in the "AFM100" region of the phase diagram predicted in Ref. [6], similar to many $4d^3$ and $5d^3$ double perovskites [8, 10-13]. The predicted influence of SOC on the $d^2$ state would inherently infer anisotropic interactions on the $Os^{6+}$ ions, with the strong Os−O hybridization in $Sr_2MgOsO_6$ ensuring that the anisotropy has significant influence on the collective properties. We have confirmed anisotropy is present in $Sr_2MgOsO_6$ via observation of the spin gap in the magnetic excitation spectrum, Fig. 7a.

The moment observed by neutron diffraction of 0.60(2) $\mu_B$ is considerably reduced (70%) from the expected high-field spin-only value of 2 $\mu_B$. Covalency has been shown to play a significant role in the reduction of the moment in the case of $4d^3$ and $5d^3$ transition metal oxides, resulting in 37-47% reductions of the $5d^3$ $Os^{5+}$ 3 $\mu_B$ moment to 1.6 to 1.9 $\mu_B$ [8-13], and we expect a similar effect in the $Os^{6+}$ $5d^2$ case. Here, however, the magnetization and specific heat analyses strongly suggests that an orbital contribution is also important, consistent with the recent x-ray magnetic circular dichroism (XMCD) study of $Os^{6+}$ in the related material $Ca_2CoOsO_6$ [48]. Our DFT results confirm that both SOC and covalency together cause the major reduction in the observed spin-moment, predicting a 47% reduction from covalency and a further 25% reduction from SOC, consistent with the total 70% reduction we observed experimentally. The result is a moment which is challenging to observe with standard neutron diffraction instrumentation, but via a high-flux, low-background experiment we were able to determine both the ground state and the moment size - it would be interesting to revisit $Ba_2CaOsO_6$ on a similar instrument in order to conclusively place the magnetic ground state among those known and predicted.

For $5d^2$ double perovskites the type I AFM structure was anticipated via theory including strong SOC in Ref. [6]. The predicted influence of SOC on the $d^2$ state would inherently infer anisotropy, with the strong Os−O hybridization in $Sr_2MgOsO_6$ ensuring that the anisotropy has significant influence on the collective properties. Via inelastic neutron scattering we have indeed observed anisotropy in $Sr_2MgOsO_6$ via the spin gap.

It is interesting to investigate the comparison between $5d^2$ and $5d^3$ systems. $Sr_2MgOsO_6$ orders at a higher temperature than other double perovskites with a single magnetic ion [28]. While $4d^3$ and $5d^3$ double perovskites like $Sr_2ScOsO_6$ have larger magnetic moments, which should yield stronger interactions and higher ordering temperatures, they also have significantly larger frustration indices ranging from 4 to 14 [8, 10-13] in comparison to 2.5 in $Sr_2MgOsO_6$. The similarity of the $QE$-space dependence and the intensity of the excitation spectra of $Sr_2MgOsO_6$ and $Sr_2ScOsO_6$ at 6 K, shown in Fig. 7, indicates that similar interaction mechanisms are responsible for the collective properties in each. However, Fig. 8a shows that the intensity of fluctuations in $Sr_2ScOsO_6$ above $T_N$ is far greater than the intensity in $Sr_2MgOsO_6$ above $T_N$. This implies strong correlations persist in $Sr_2ScOsO_6$ in the absence of long range magnetic order – a hallmark of frustration. While the tetragonal symmetry of $Sr_2MgOsO_6$ may play some role in reducing the geometric frustration, it is less distorted than monoclinic $Sr_2ScOsO_6$, which has angles that deviate more substantially from 180° [8].

The question, therefore, is why is the frustration relieved in $Sr_2MgOsO_6$ compared to $Sr_2ScOsO_6$? The strength of hybridization plays a major role in determining the strength of the interactions, but does not directly explain relief of frustration. Similarly, the unit cell volume and B/B′ site disorder affect interaction strengths, and all these mechanisms likely contribute to differences in $T_N$s amongst the AFM $d^2$ DPs $Sr_2MgOsO_6$ ($|\Theta|/T_N = 2.5$) [this work], $Ba_2CaOsO_6$ ($|\Theta|/T_N = 3.1$) [51] and $Ca_2CaOsO_6$ ($|\Theta|/T_N = 3.0$) [52], but in all of these materials the frustration index is low compared to $d^3$ DP $Sr_2ScOsO_6$. Our DFT results reveal that the SOC induced anisotropy leads to selection of the magnetic ground state, as also found experimentally for

Sr$_2$ScOsO$_6$ [38], therefore stronger anisotropy promotes a more robust ground state. Given the observation of the large spin gap in Sr$_2$MgOsO$_6$ and the multiple sources of evidence we have presented for stronger SOC in this $d^2$ material, we conclude that SOC induced anisotropy is the dominant factor in the relieved frustration in Sr$_2$MgOsO$_6$.

This conclusion is supported by an earlier theoretical prediction that $T_N$ should vary with the strength of anisotropy in such systems, due to reduced competition between possible ground states [53]. The anisotropy term is proportional to the gap size (probed by INS) divided by the moment size (probed by neutron diffraction). Given that the gap is of the same magnitude in Sr$_2$MgOsO$_6$ and Sr$_2$ScOsO$_6$, Fig. 7, but the moment is less than half the size in Sr$_2$MgOsO$_6$, we conclude that the anisotropy is indeed much larger in Sr$_2$MgOsO$_6$. Therefore, the results presented demonstrate the relief of frustration via SOC induced anisotropy, and represent the experimental demonstration of the evolution of $T_N$ with SOC.

## Conclusion

The double perovskite osmate Sr$_2$MgOsO$_6$ has been synthesized and characterized by magnetometry, specific heat measurements, and elastic and inelastic neutron scattering. The combined results demonstrate that spin-orbit coupling is essential to describe the magnetic properties of the system. The Os$^{6+}$ moments order antiferromagnetically at 108 K in a type I configuration on the Os fcc sublattice as theoretically predicted [6]. For the first time in such a 5$d^2$ material, refinements of neutron powder diffraction data yields a moment of 0.60(2) µ$_B$ per Os cation which is significantly reduced due to the combined effects of SOC and covalency. Through comparison of inelastic neutron spectra, it is shown that SOC induced anisotropy has the ability to relieve frustration in $d^2$ systems, relative to analogous $d^3$ materials which have systematically higher frustration indices, promoting high magnetic transition temperatures.

# Author Information

**Corresponding Author**

r.c.morrow@ifw-dresden.de

**Notes**

The authors declare no competing financial interest.

**Author Contributions**

RM synthesized all samples as well as conducted and analyzed XRD and POWGEN NPD data. RM and JX collected and analyzed magnetometry data. SR, AUBW, SW, and BB were responsible for the specific heat data collection and analysis. AET and AAA conducted the HB-1A NPD experiment. AET, MBS, and AIK measured the SEQUOIA INS data. AET analyzed the results from the HB-1A and SEQUOIA experiments. ADC supervised neutron scattering activities, and PMW supervised synthesis and property characterization activities. RM and AET led the manuscript preparation, and all coauthors contributed.

# Acknowledgements

Support for this research was provided by the Center for Emergent Materials an NSF Materials Research Science and Engineering Center (DMR-1420451), and in the framework of the materials world network (Deutsche Forschungsgemeinschaft DFG project no. WU595/5-1 and National Science Foundation (DMR-1107637)). S. Wurmehl gratefully acknowledges funding by DFG in project WU 595/3-3 (Emmy Noether program) and by DFG in SFB 1143. A portion of this research was carried out at Oak Ridge National Laboratory's Spallation Neutron Source and High Flux Isotope Reactor, which is sponsored by the U.S. Department of Energy, Office of Basic Energy Sciences. Work at the University of Missouri (DJS) was funded through the Department of Energy S3TEC Energy Frontier Research Center, award DE-SC0001299/DE-FG02-09ER46577. The authors would like to acknowledge S. Calder and M. D. Lumsden for helpful discussions, and the authors also thankfully acknowledge Ashfia Huq for experimental assistance with POWGEN data collection.

This manuscript has been authored by UT-Battelle, LLC under Contract No. DE-AC05-00OR22725 with the U.S. Department of Energy. The United States Government retains and the publisher, by accepting the article for publication, acknowledges that the United States Government retains a non-exclusive, paid-up, irrevocable, worldwide license to publish or reproduce the published form of this manuscript, or allow others to do so, for United States Government purposes. The Department of Energy will provide public access to these results of federally sponsored research in accordance with the DOE Public Access Plan (http://energy.gov/downloads/doe-public-access-plan).

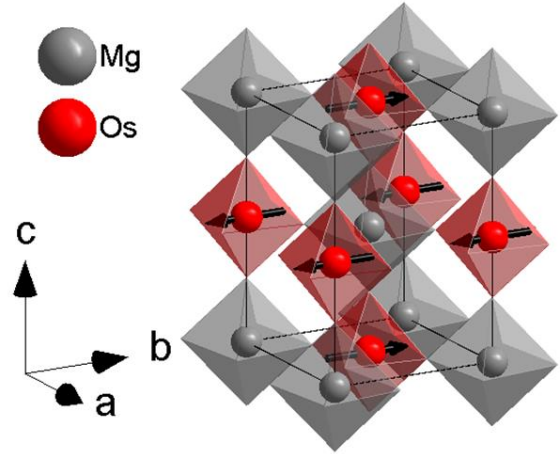

FIG. 1 (color online). The crystal and magnetic structure of $Sr_2MgOsO_6$ with Mg and Os shown as grey and red spheres located within octahedra of the same color with O ions positioned at the corners. Sr cations are omitted for clarity. The magnetic moment on $Os^{6+}$ is shown as a black arrow, and while shown along the $b$ axis, is known only to be in the $a$-$b$ plane.

| Temperature (K) | 10 | 50 | 300 |
|---|---|---|---|
| Space Group | $I4/m$ | $I4/m$ | $I4/m$ |
| $a$ (Å) | 5.52776(8) | 5.52889(7) | 5.56345(6) |
| $c$ (Å) | 7.9781(1) | 7.9755(1) | 7.9217(1) |
| $V$ (Å)$^3$ | 243.781(9) | 243.798(8) | 245.193(7) |
| $R_{wp}$ | 3.25% | 3.43% | 3.47% |
| $R_p$ | 2.00% | 2.30% | 2.22% |
| | | | |
| Mg−O1 (×4, Å) | 2.037(1) | 2.039(1) | 2.041(1) |
| Mg−O2 (×2, Å) | 2.060(2) | 2.059(2) | 2.047(2) |
| Os−O1 (×4, Å) | 1.904(1) | 1.903(1) | 1.911(1) |
| Os−O2 (×2, Å) | 1.929(2) | 1.928(2) | 1.914(2) |
| ∠Mg−O1−Os (°) | 165.32(4) | 165.33(4) | 169.01(6) |
| | | | |
| O1 $x$ | 0.2264(2) | 0.2266(2) | 0.2343(2) |
| O1 y | 0.2907(2) | 0.2909(2) | 0.2824(2) |
| O2 z | 0.2582(2) | 0.2582(2) | 0.2584(2) |
| | | | |
| Sr $U_{iso}$ | 0.0023(1) | 0.0025(1) | 0.0079(1) |
| Mg = Os $U_{iso}$ | 0.0013(1) | 0.0014(1) | 0.0036(1) |
| O1 $U_{eq}$ | 0.0043(1) | 0.0044(1) | 0.0098(1) |
| O2 $U_{eq}$ | 0.0043(1) | 0.0046(2) | 0.0115(2) |

TABLE 1: Neutron powder diffraction parameters obtained from Rietveld refinement for $Sr_2MgOsO_6$ at 10, 50 and 300 K. Mg and Os $U_{iso}$ values were constrained to be equal. $U_{eq}$ is calculated as a third of the trace of the tensor for the anisotropically refined oxygen ion U's.

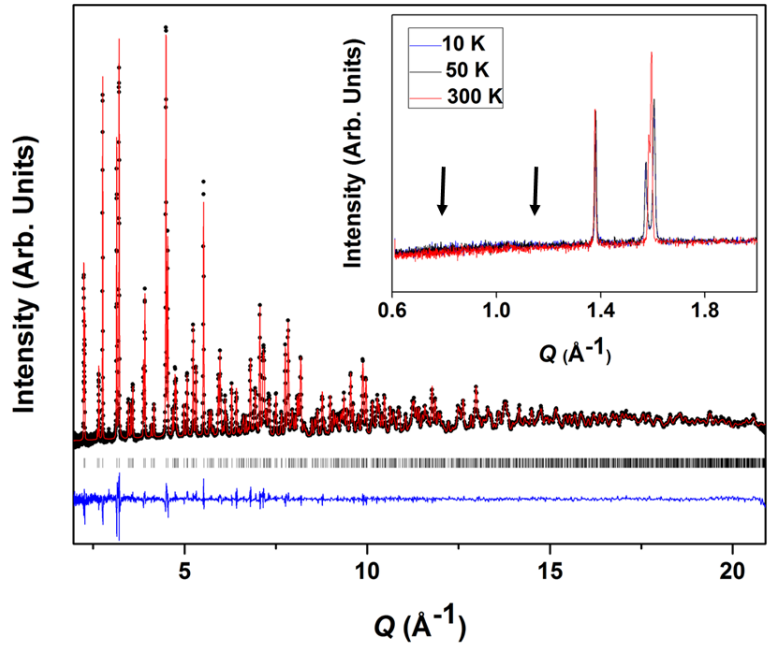

FIG. 2 (color online). Refined neutron powder diffraction pattern of $Sr_2MgOsO_6$ at 10 K. Black symbols, red curves, and blue curves correspond to the observed data, the calculated patterns, and the difference curve, respectively. The black hashes correspond to the nuclear peak positions of the compound. The inset shows low $Q$ neutron powder diffraction data on $Sr_2MgOsO_6$ from POWGEN at 300, 50, and 10 K. Arrows indicate the position of the potential magnetic reflections associated with Type I AFM order.

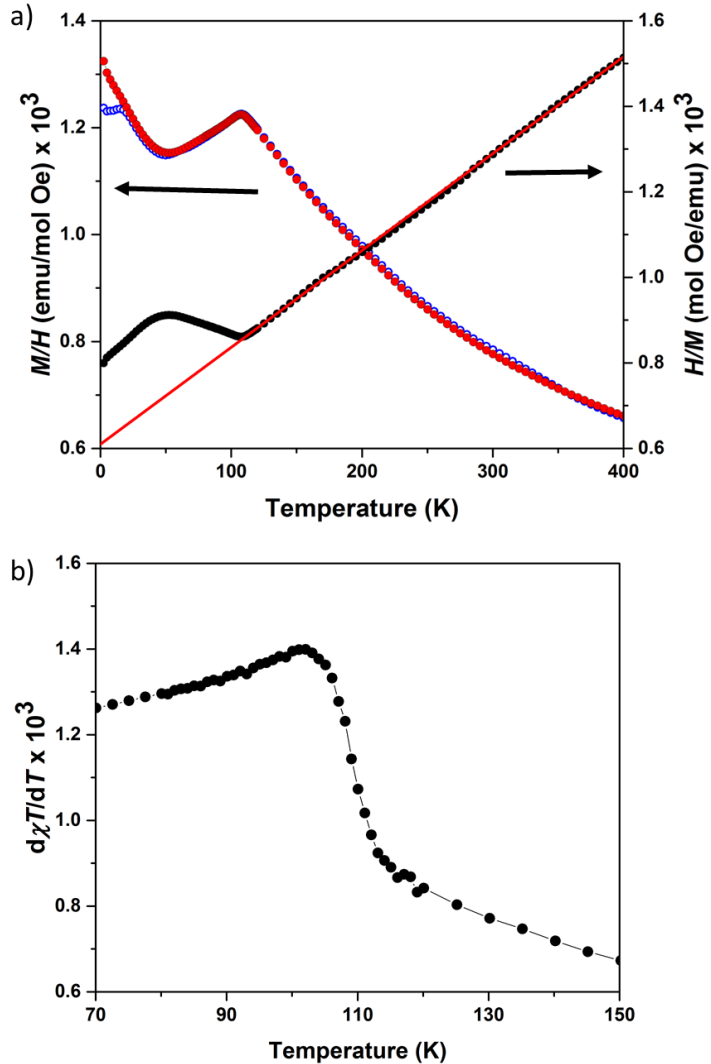

FIG. 3 (color online). a) The temperature dependence the field-cooled (red filled circles) and zero-field-cooled (blue open circles) magnetization under an applied field of $Sr_2MgOsO_6$ plotted against the left axis. The inverse data (black circles) is plotted against the right axis with a red line indicating the higher temperature Curie-Weiss fitting. b) The Fisher heat capacity of $Sr_2MgOsO_6$ derived from the data in panel a).

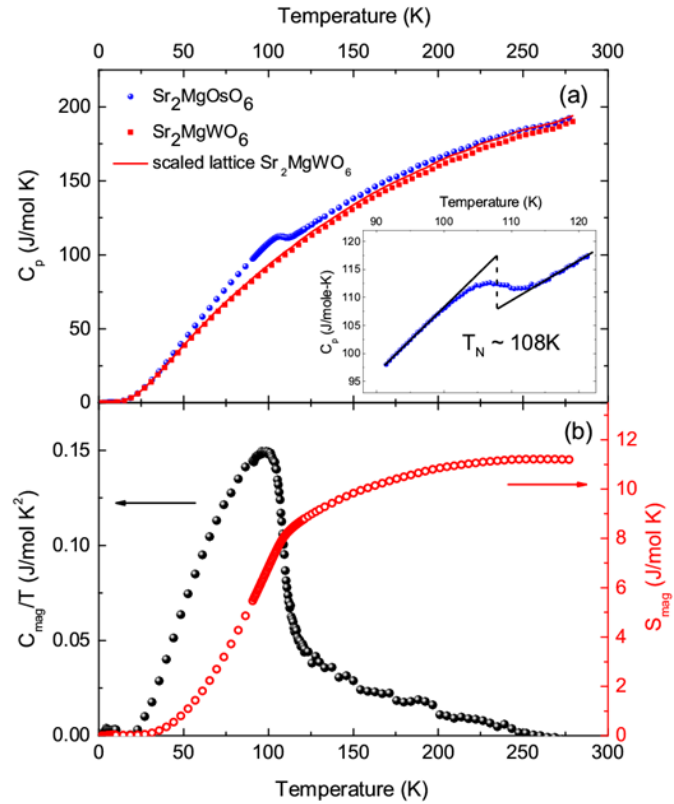

FIG. 4 (color online). (a) The temperature dependence of the specific heat of $Sr_2MgOsO_6$ (blue) and diamagnetic $Sr_2MgWO_6$ (red) which is used to approximate the lattice specific heat of $S_2MgOsO_6$. (b) The magnetic specific heat (black) of $Sr_2MgOsO_6$ taken by subtracting the scaled specific heat of $Sr_2MgWO_6$ and magnetic entropy (red) obtained by integration.

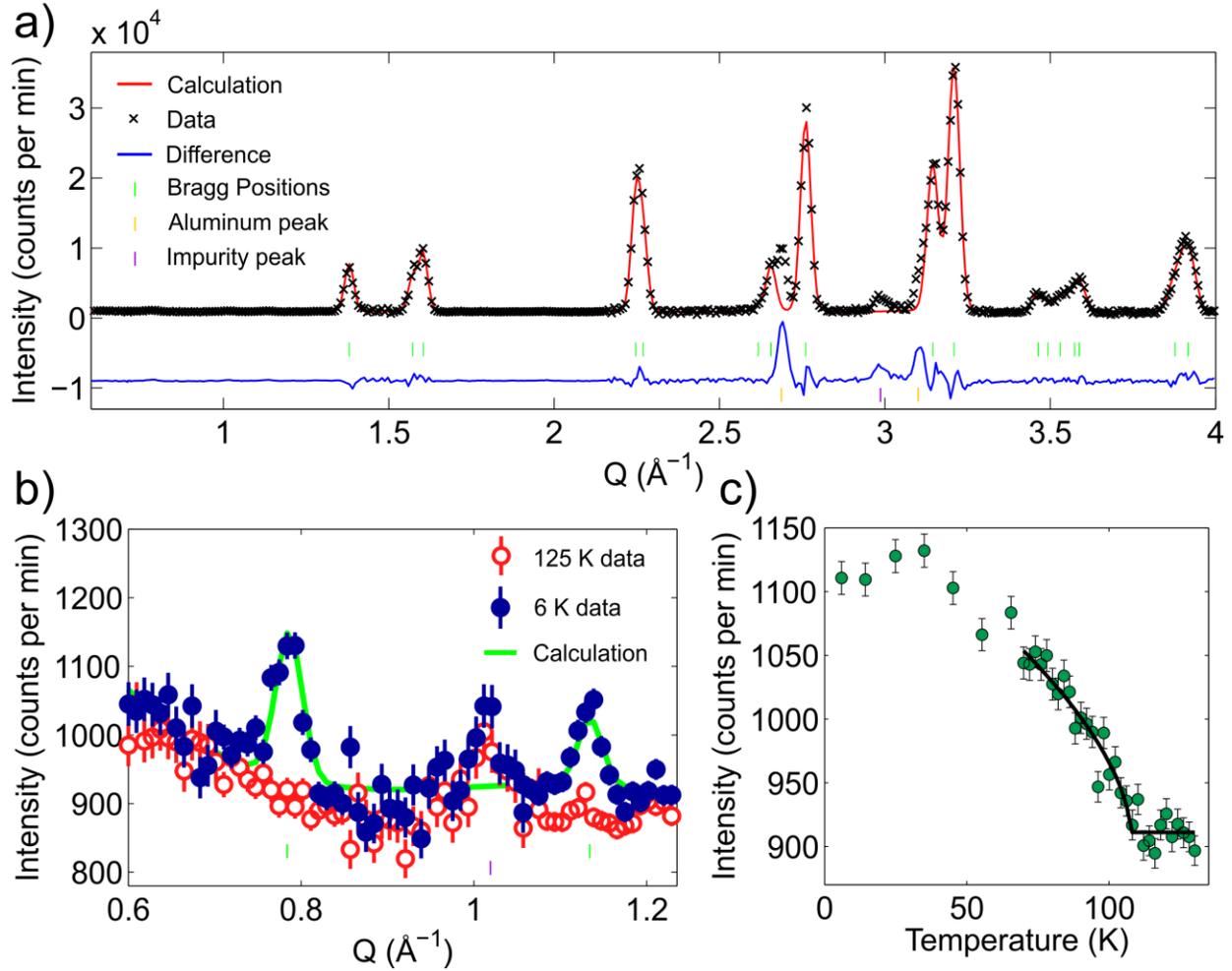

FIG. 5 (color online). a) Neutron powder diffraction data collected on $Sr_2MgOsO_6$ using the HB-1A beamline at 125 K. b) A close view of the low $Q$ data at 125 K (red open circles) and 6 K (blue filled circles) with a Rietveld fitting to the 6 K data (green line) as described in the text. c) The temperature dependence of the $Q(001) = 0.78$ Å$^{-1}$ magnetic reflection with a power-law fitting to the data as described in the text.

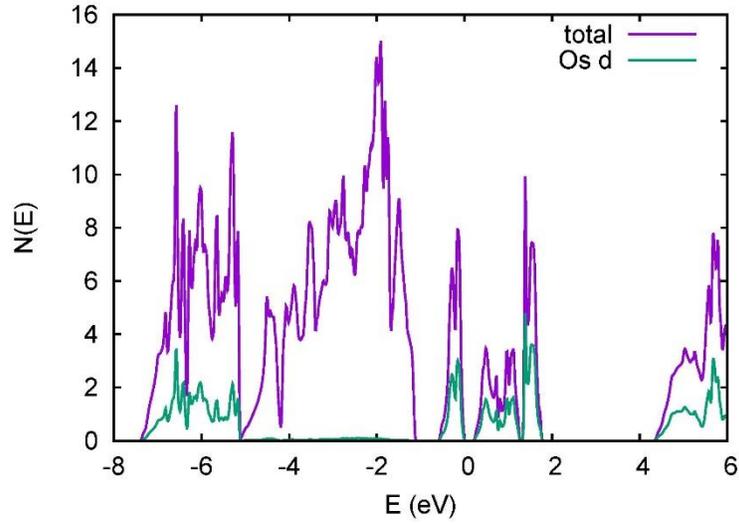

FIG. 6 (color online) PBE+U (U=3 eV) density of states and projection onto Os for the ground state antiferromagnetic order, including SOC.

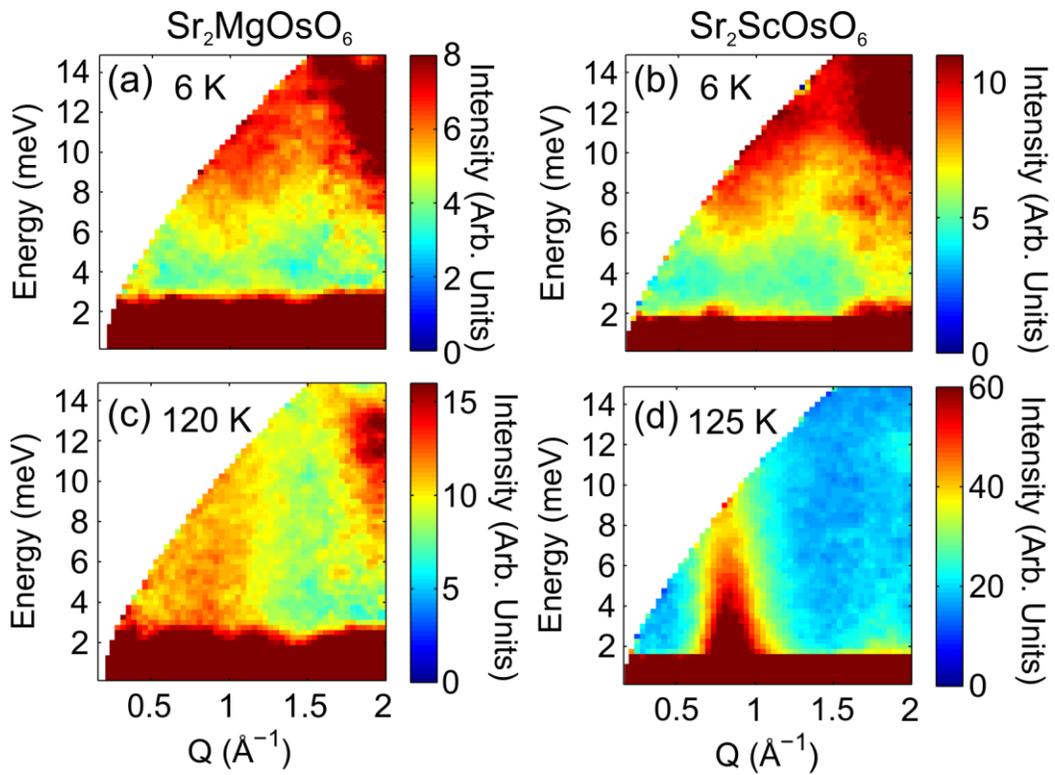

FIG. 7 (color online). Neutron scattering intensity maps, measured with $E_i = 20$ meV, showing the gap below $T_N$ for (a) $Sr_2MgOsO_6$ and (b) $Sr_2ScOsO_6$ which closes above $T_N$ in (c) and (d) respectively.

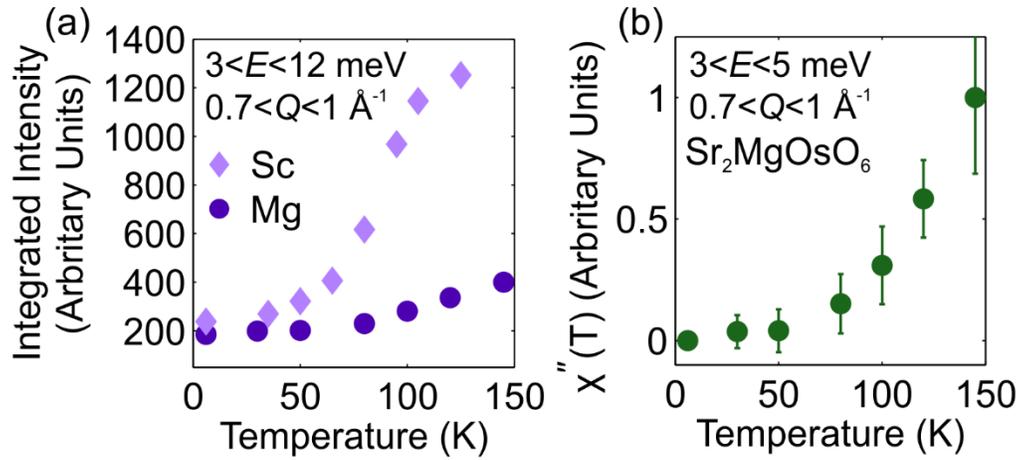

FIG. 8 (color online). (a) Comparison of the evolution of normalized per Os integrated neutron scattering intensity of $Sr_2MgOsO_6$ and $Sr_2ScOsO_6$ and (b) the temperature dependence of the scattering within the gap of $Sr_2MgOsO_6$, converted to $\chi''$ as in Ref. [10].